\documentclass[aps,twocolumn,prl,superscriptaddress]{revtex4-1}
\usepackage{amsmath}
\usepackage{amssymb}
\usepackage[final]{graphicx}% Include figure files
\usepackage{dcolumn}% Align table columns on decimal point
\usepackage{bm}% bold math
\usepackage{color}
\usepackage{setspace}

\begin{document}

\title{Quantifying material properties of cell monolayers by analyzing integer topological defects}

\author{Carles Blanch-Mercader}
\email{co-first author}
\affiliation{Department of Biochemistry, University of Geneva, 1211 Geneva, Switzerland}
\affiliation{Department of Theoretical Physics, University of Geneva, 1211 Geneva, Switzerland}
 
\author{Pau Guillamat}
\email{co-first author}
\affiliation{Department of Biochemistry, University of Geneva, 1211 Geneva, Switzerland}

\author{Aur\'elien Roux}
\email{aurelien.roux@unige.ch}
\affiliation{Department of Biochemistry, University of Geneva, 1211 Geneva, Switzerland}
\affiliation{NCCR Chemical Biology, University of Geneva, 1211 Geneva, Switzerland}

\author{Karsten Kruse}
\email{karsten.kruse@unige.ch}
\affiliation{Department of Biochemistry, University of Geneva, 1211 Geneva, Switzerland}
\affiliation{Department of Theoretical Physics, University of Geneva, 1211 Geneva, Switzerland}
\affiliation{NCCR Chemical Biology, University of Geneva, 1211 Geneva, Switzerland}

\begin{abstract}
In developing organisms, internal cellular processes generate mechanical stresses at the tissue scale. The resulting deformations depend on the material properties of the tissue, which can exhibit long-ranged orientational order and topological defects. It remains a challenge to determine these properties on the time scales relevant for developmental processes. Here, we build on the physics of liquid crystals to determine material parameters of cell monolayers. Specifically, we use a hydrodynamic description to characterize the stationary states of compressible active polar fluids around defects. We illustrate our approach by analyzing monolayers of C2C12 cells in small circular confinements, where they form a single topological defect with integer charge. We find that such monolayers exert compressive stresses at the defect centers, where localized cell differentiation and formation of three-dimensional shapes is observed.
\end{abstract}
  
\maketitle

Mechanics is essential for understanding morphogenesis in developing organisms. Tissues are formed by collections of cells, which move and deform through the action of energy-consuming biochemical reactions. Increasing efforts combine experiments and theory to determine the physical conditions of tissue morphogenesis~\cite{Behrndt:2012gy,Etournay:2015cn,Doubrovinski:2017if,Mongera:2018ka}. These include material properties like elasticity and viscosity as well as active stresses, that is, stresses generated by burning a chemical fuel, which often is Adenosine-Triphosphate (ATP). However, in a developing organism the situation is complicated by regulatory processes on various levels, such that it is often difficult to disentangle mechanics and chemistry. 

In an effort to reduce this mutual dependence, experiments on collections of cells in well-controlled environments have been developed~\cite{Ladoux:2017fe,Hakim:2017je,RocaCusachs:2017ge}. However, direct measurements of material parameters are rarely feasible on the time scales associated with developing organisms, which range from hours to days and longer. Alternatively, material parameters of tissues can be extracted from in vitro experiments by using theoretical descriptions~\cite{Delarue:2013ck,Nier:2016fb,Duclos:2018ita,Tlili:2018di,Wyatt:2020ht,Fouchard:2020fz,Trushko:2019ko}. Two major theoretical approaches have been used in this context: on one hand, vertex models, which generalize the descriptions of foams to account for cellular processes~\cite{Kaefer:2007do,Farhadifar:2007hr,HocevarBrezavscek:2012ji,Park:2015ih}; on the other hand, continuum descriptions, which focus on coarse-grained properties at the tissue level~\cite{Goriely:2005iq,Bittig:2008ku,Hannezo:2011ko,Douezan:2011is,Lee:2011iz}. In spite of the progress made, a complete characterization of the material properties of tissues is still lacking.

This holds, in particular, when tissues exhibit orientational order, which results form anisotropies at the cell level. In this way, tissues acquire properties of liquid crystals~\cite{Gruler:1999bt,Duclos:2014bs}. Similar to these materials, tissues can exhibit topological defects, where  orientation is ill-defined~\cite{deGennes:2002vq}. Recent experiments show that the position of defects correlates with biological processes essential for development~\cite{Saw:2017gn,Kawaguchi:2017em,MaroudasSacks:2020eh}. In the context of liquid crystals, topological defects have been used to determine material parameters associated with the orientational degrees of freedom~\cite{Hudson:1989dd,Brugues:2008hx}. In this work, we build on these ideas for characterizing mechanical properties of cell monolayers. 

\begin{figure}[b] %  figure placement: here, top, bottom, or page
\centering
\includegraphics[width=8.6cm]{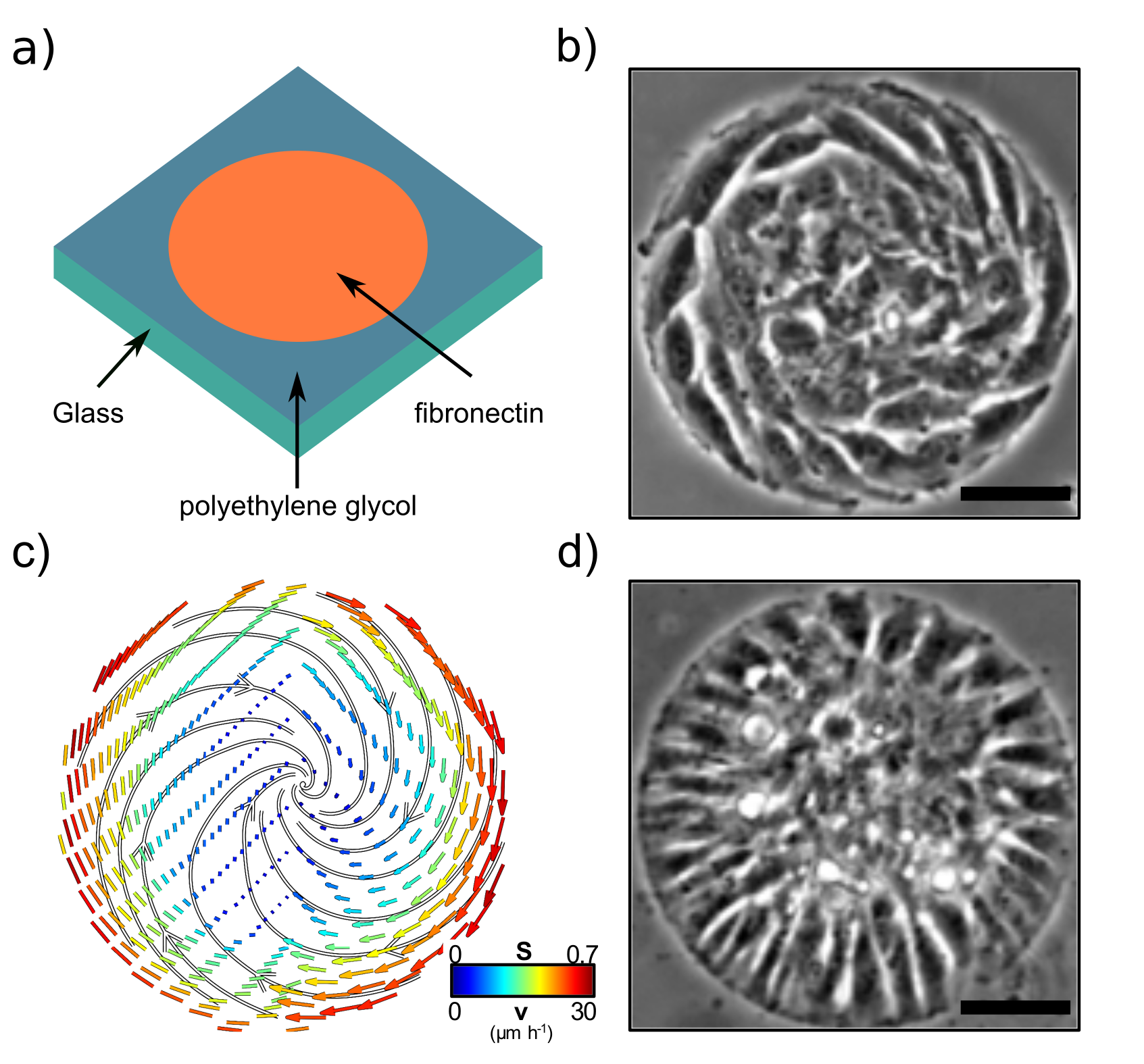}
\caption{(color online) Confined C2C12 monolayers. a) Schematic of the experimental setup. b) Phase-contract image of a spiral in a circular domain of 100~$\mu$m radius. c) Orientational order (left) and velocity fields (right) averaged over $N=12$ spirals. Colors correspond to the polar order parameter $S$ and speeds, see legend. Gray lines: velocity stream lines. d) Phase-contrast image of an aster in a circular domain of 100~$\mu$m radius. Scale bar in (b,d): 50~$\mu$m.}\label{fig1}
\end{figure}
We confined C2C12 myoblasts to fibronectin-coated circular domains with radii between 50~$\mu$m and 150~$\mu$m that are surrounded by a non-adhesive surface coated with polyethylene glycol (PEG), Fig.~\ref{fig1}a. In the course of the experiment, the cell number increased by proliferation. For all the time points discussed in this work, the cells formed a monolayer. Upon reaching confluence, when cells fully cover a domain, the myoblasts arranged into spirals, Fig.~\ref{fig1}b. In this state, cells approximately aligned with the domain boundary and exhibited collective rotation around its center, Fig.~\ref{fig1}c. In contrast to some epithelial monolayers~\cite{Doxzen:2013hx}, the presence of shear flows in the azimuthal direction, see Fig.~S1, suggested that our system behaved as a viscous fluid on long time scales. Approximately 20~h after having reached confluence, the spiral changed into an aster, where cells at the periphery oriented perpendicularly to the boundary, Fig.~\ref{fig1}d. No collective rotation was observed then. In both asters and spirals, the cells were disorganized at the center, Fig.~\ref{fig1}b,d. Furthermore, both cell arrangements were on average rotationally invariant. For details on the experimental setup see Ref.~\cite{Guillamat:2020te}

To describe the organization of C2C12 monolayers, we use the polarization field $\mathbf{p}$ with the polar order parameter $S=|\mathbf{p}|$ that determines the local degree of orientational order similar to liquid crystals~\cite{deGennes:2002vq}. Topological defects of $\mathbf{p}$ are characterized by their charge. It is given by the number of turns of $\mathbf{p}$ along a closed trajectory around the defect center, where $S=0$. Asters and spirals are two examples of defects with charge +1. 

In the following, we use a continuum approach to study integer topological defects of cell monolayers. In addition to $\mathbf{p}$, we introduce the cell number density $n$ and the velocity field $\mathbf{v}$. Since the time scale associated with the shear flows~\cite{Duclos:2018ita} is much smaller than the proliferation time, we consider that the cell number is conserved. In polar coordinates $r$ and $\theta$ and assuming rotational invariance as observed experimentally, we have in steady state
\begin{align}
\partial_{r}(n v_{r})+\frac{n v_r}{r}&=0.\label{eq:massbalanceprimer}
\end{align}
As there is no influx of cells into the circular domain, that is, $v_r(r=R)=0$, we have $v_r(r)=0$ for $0\le r\le R$.

The typical scales associated with cell flows imply a Reynolds number $Re\ll1$. As a consequence, the conservation of momentum reduces to force balance, which in polar coordinates reads
\begin{align}
\partial_{r}\sigma_{rr}^\mathrm{tot}+\frac{\sigma_{rr}^\mathrm{tot}-\sigma_{\theta\theta}^\mathrm{tot}}{r}&=-T_0 p_r \label{eq:forcebalance1}\\
\partial_{r}\sigma_{\theta r}^\mathrm{tot}+\frac{\sigma_{\theta r}^\mathrm{tot}+\sigma_{r\theta}^\mathrm{tot}}{r}&=\xi v_{\theta}-T_0 p_\theta.\label{eq:forcebalance2}
\end{align}
The divergence of the total stress tensor $\mathsf{\sigma}^\mathrm{tot}$ on the left-hand side is balanced by the monolayer-substrate interaction forces on the right-hand side. The latter are composed of viscous friction between the monolayer and the substrate, which is characterized by the effective friction coefficient $\xi$, and an active polar traction force with the scale set by $T_0$. Even though traction force appears as an external force in Eqs.~\eqref{eq:forcebalance1} and \eqref{eq:forcebalance2}, it results from cellular activity.

We can express the total stress in terms of the velocity and polarization fields for an active polar fluid by following the standard procedure of non-equilibrium thermodynamics~\cite{Kruse:2004il,Furthauer:2012iu}. The total stress can be decomposed into the Ericksen stress $\sigma^e$ and a deviatory part $\sigma$. The Ericksen stress is a generalization of the hydrostatic pressure for polar fluids~\cite{deGennes:2002vq}. For the deviatory part, we find
\begin{align}
\sigma_{rr,\theta\theta}&=\mp\frac{1}{2}S^2 \cos(2\psi)\zeta\Delta\mu-S^2\zeta''\Delta\mu  \nonumber\\
&\pm\frac{\nu}{2}S\left(h_\parallel \cos(2\psi)-h_\perp \sin(2\psi)\right)   +\nu' S h_\parallel \label{eq:devStressTensorRR} \\
\sigma_{r\theta,\theta r}&=2\eta v_{r\theta}-\frac{1}{2}S^2 \sin(2\psi)\zeta\Delta\mu  \nonumber\\
&+\frac{\nu}{2}S\left(h_\parallel \sin(2\psi)+h_\perp \cos(2\psi)\right) \pm\frac{S h_\perp}{2},\label{eq:devStressTensorRTheta} 
\end{align}
where the upper (lower) signs correspond to the first (second) index pair. 

In these equations, the polarization field $\mathbf{p}$ is expressed through the polar order parameter $S$ and the angle $\psi$ with respect to the radial direction. The molecular field $\mathbf{h}$ is related to the system's free energy $\mathcal{F}$ by $\mathbf{h}=-\delta\mathcal{F}/\delta \mathbf{p}$. Its components perpendicular and parallel to the direction of $\mathbf{p}$ are $h_\perp$ and $h_\parallel$. The non-vanishing components of the velocity gradient tensor are $v_{\theta r}=v_{r\theta}=(\partial_r v_\theta-v_\theta/r)/2$ and $\eta$ is the shear viscosity. The mechanical stress is coupled to active processes by the constants $\zeta$ and $\zeta''$, where $\Delta\mu$ is the chemical potential difference that drives ATP hydrolysis. Finally, the flow alignment parameters $\nu$ and $\nu'$ couple the stress to $\mathbf{h}$. For details of the derivation, see Ref.~[LongArticle].

The corotational convective time derivative $\frac{D}{Dt}\mathbf{p}$ reads in steady state
\begin{align}
\frac{h_\parallel}{\gamma}-\nu S v_{r\theta}\sin(2\psi)&=0\label{eq:dinamicadirectorparticular1}\\
\frac{h_\perp}{\gamma}+S v_{r\theta}\left(1-\nu \cos(2\psi)\right)&=0 .\label{eq:dinamicadirectorparticular2}
\end{align}
Here, the rotational viscosity $\gamma$ controls the dissipation related to reorientation of $\mathbf{p}$. The coupling coefficient $\nu$ between $\frac{D}{Dt}\mathbf{p}$ and the velocity gradient tensor is imposed by an Onsager relation. Since we assume that $\Delta\mu$ is constant, we do not give the expression for the ATP-hydrolysis rate. 

The free energy $\mathcal{F}=2\pi\int_0^R f r dr$, where $f$ is the free-energy density, penalizes deviations of the cell number density from a reference density $n_0$ and deviations of $\mathbf{p}$ from the disordered state with $S=0$. In addition, spatial distortions of $\mathbf{p}$ are accounted for by the Frank free energy in the one-constant approximation~\cite{deGennes:2002vq}. Explicitly,
\begin{align}
f&=\!\frac{B}{2}\!\left(\frac{n}{n_0}-1\right)^2\!\!+\frac{\chi}{2}S^2\!+\frac{{\cal K}}{2}\left[(\partial_r S)^2\!+S^2(\partial_r\psi)^2\!+\frac{S^2}{r^2}\right]\nonumber
\end{align}
with $B$, $\chi$, and $\mathcal{K}$ being the respective elastic moduli. By analyzing the organization of $\mathbf{p}$ around semi-integer topological defects, we found that the one-constant approximation is appropriate for C2C12 monolayers~[LongArticle].

It remains to fix the boundary conditions. Due to rotational invariance of our system and to obtain regular solutions, we have $v_\theta(r=0)=0$. At $r=R$, we consider the absence of mechanical stresses in the azimuthal direction, $\sigma_{\theta r}^\mathrm{tot}(r=R)=0$. We focus on defect configurations and thus assume $S=0$ in the center and to be finite at the domain boundary. Without restrictions of generality, we choose $S(r=R)=1$. For the angle $\psi$, we impose $\partial_r\psi|_{r=0,R}=0$. 

In the following, we apply this framework to topological defects in confined myoblast monolayers and thus determine their material parameters. A uniform polarization angle $\psi=\psi_0$ is a solution to Eq.~\eqref{eq:dinamicadirectorparticular2} and is compatible with the boundary conditions. The flow alignment parameter $\nu$ can be obtained from this polarization angle through the relation $\nu\cos(2\psi_0)=1$, see Eq.~\eqref{eq:dinamicadirectorparticular2}. In our experiments, we determined the polarization angle by a tensor structure method~\cite{Puspoki:2016um,Guillamat:2020te}[LongArticle] for a domain with radius 100~$\mu$m~\cite{Guillamat:2020te}. For C2C12 monolayers, we found $\nu=-1.1\pm0.3$ (mean$\pm$std, N=12 domains) corresponding to $\psi_0\lesssim 90^\circ$. In passive liquid crystals, such a value corresponds to rod-shaped molecules, which is compatible with the elongated shape of C2C12 myoblasts. Note that our value is also in the range of values determined in epithelial monolayers of fly wing discs~\cite{Aigouy:2010fl}.

\begin{figure}[t] %  figure placement: here, top, bottom, or page
\centering
\includegraphics[]{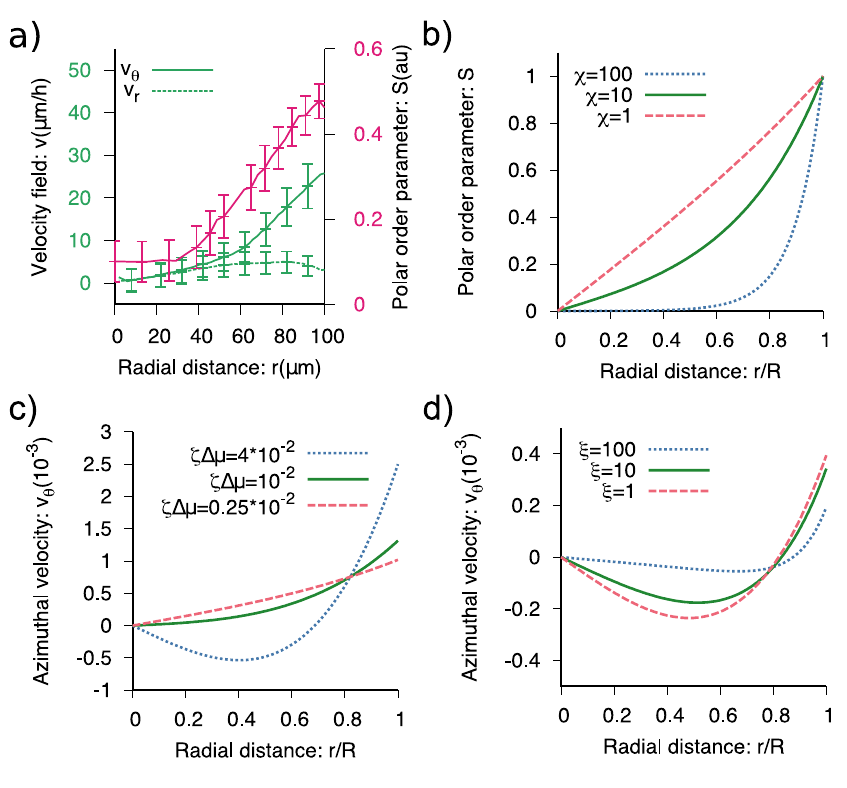}
\caption{(color online) Spiral steady-state patterns. a) Profiles of the polar order parameter $S$ and the velocity components $v_r$ and $v_\theta$ for myoblast monolayers (mean$\pm$sem, $N=12$). b) Theoretical profiles of $S$ for varying values of $\chi$. c,d) Theoretical profiles of $v_\theta$ for varying $\zeta\Delta\mu$ (c) and varying $\xi$ (d). Unless stated otherwise parameters are $\nu=-1.2$, $\zeta\Delta\mu=10^{-2}$, $\eta=\xi=\chi=1$, $T_0=10^{-3}$ in (b,c) and $T_0=0$ in (d). The units are set by $\mathcal{K}=\gamma=R=1$.}\label{fig2}
\end{figure}

To determine the remaining mechanical parameters, we fit the steady-state solutions of our system to the polar order $S$ and azimuthal velocity $v_\theta$ in spirals for domains with radii $R=50~\mu$m, 100~$\mu$m, and 150~$\mu$m, Fig.~\ref{fig2}a. The velocity field was measured through particle image velocimetry~\cite{Guillamat:2020te}. To perform the fits, we first non-dimensionalize the equations by introducing the length scale $R$, the velocity scale $\mathcal{K}/R\gamma$, and the energy scale $\mathcal{K}$. 

Examples of numerical steady-state solutions to our equations are given in Fig.~\ref{fig2}b-d. The value of $S$ increases monotonically from the center to the boundary, Fig.~\ref{fig2}b. When flow gradients are small, i.e., $|\gamma\nu v_{r\theta}\sin(2\psi_0)|\ll\chi$, then the polar order is induced by the boundaries in a layer of size $\sqrt{\mathcal{K}/\chi}$~[LongArticle]. 

The velocity profile depends on the polarization field, but is also determined by the competition between the different active and dissipative terms~[LongArticle]. For fixed parameter values of the dissipative terms, the flow patterns change from a coherent rotation to counterpropagating flows as the ratio between traction forces and the active stresses is decreased, see Fig.~\ref{fig2}c. Note that in all cases the net torque on the system vanishes. Flows that are driven by gradients in the active stress decay from the boundary over a length scale $\ell^2=(\eta+\gamma\tan(2\psi_0)^2/4)/\xi$, see Fig.~\ref{fig2}d. 

The goodness of our fits is quantified by an error function defined as
\begin{align}
\mathcal{E}&=\sum_i |v_{\theta,i}^{e}-v_{\theta,i}|\Delta r_i+\sum_i|S_i^{e}-S_i| \Delta r_i.\label{eq:errorfunction}
\end{align}
Here, the superscript 'e' indicates values averaged over at least $N=5$ experiments, and the index $i$ indicates that samples are taken at discrete radial positions $r_i$. Furthermore, $\Delta r_i=r_{i+1}-r_i$ is related to the experimental spatial resolution and $\Delta r_i\sim 5~\mu$m. There is one parameter set that yields the minimal error $\mathcal{E}_\mathrm{min}$. 
\begin{table*}[t]
\begin{ruledtabular}
\begin{tabular}{cccccccc}
$T_0(\text{Pa})$ & $\zeta\Delta\mu(\text{kPa}~\mu\text{m})$ & $\eta(\text{kPa h}~\mu\text{m})$ & $\xi(\text{Pa h}/\mu\text{m})$ & $\sqrt{{\cal K}/\chi}(\mu\text{m})$ & $\nu$ & $B/n_0(\text{kPa}~\mu\text{m}^3)$ & $n^{\mathrm{tot}}(10^{-3}~\mu\text{m}^{-2})$ \\
$<600\pm60$ & $48\pm4$ & $34\pm8$ & $<40\pm20$ & $>50$ & $-1.1\pm0.3$ & $4600\pm800$ & $8.2\pm0.5$ 
  			
\end{tabular}
\end{ruledtabular}
\caption{Table of material parameters for active stress dominated solutions in Figs.~\ref{fig3}, \ref{fig4}. For converting 3d material properties into 2d material properties, we used a height of $10$~$\mu$m for the cell monolayer.  \label{tab:table1}}	
\end{table*}

\begin{figure}[b] 
\centering
\includegraphics[]{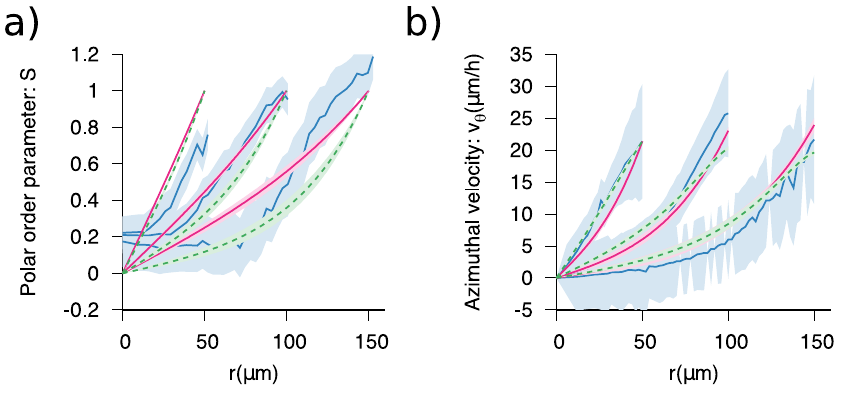}
\caption{(color online) Theoretical fits to experimental data. a) Polar order parameter $S$ and b) azimuthal velocity $v_\theta$ as a function of the radial distance $r$. Averaged experimental profiles (blue, $N=11, 12, 5$ for confining domain radius $50, 100, 150$~$\mu$m), mean fit in the active-stress dominated (magenta, full lines) and in the traction dominated parameter region (green, dashed lines). The theoretical curves are endowed with physical units such that $S(R)=1$ and $v_\theta(R)=21.4~\mu$m/h for $R=50~\mu$m. Error bars in theoretical fits correspond to std of all values of parameters with $\mathcal{E}<1.1\mathcal{E}_\mathrm{min}$, and experimental curves to sem.}\label{fig3}
 \end{figure}

Note that in the experiments shown in Fig.~\ref{fig3}, the value of the polar order parameter $S$ at the boundary depends on the radius $R$ of the domain. It is currently not clear, how this dependence arises. Furthermore, $S$ does not decay to zero in the center of the domain. This could be due to our experimental resolution. For simplicity, in our theory, we refrained from considering these effects. To account for the various sources of variability, we consider regions of parameter sets with an error $\mathcal{E}<1.1\mathcal{E}_\mathrm{min}$ to be acceptable. 

We systematically scanned the parameter space and, by applying our goodness criterion, discovered two distinct regions in parameter space that are compatible with the polarization and azimuthal velocity fields in spirals of C2C12 monolayers~[LongArticle]. The mean fits to $S$ and $v_\theta$ are shown in Fig.~\ref{fig3}. In both regions the penetration length of the polar order field is comparable to or larger than the size of the confinement $R$, which is consistent with having a single integer defect. In contrast, the solutions differ in the dominant active and dissipative mechanisms: In one region, substrate interactions dominate flow generation, whereas in the other region, gradients of active stresses are the main driving force.

\begin{figure}[b] %  figure placement: here, top, bottom, or page
\centering
\includegraphics{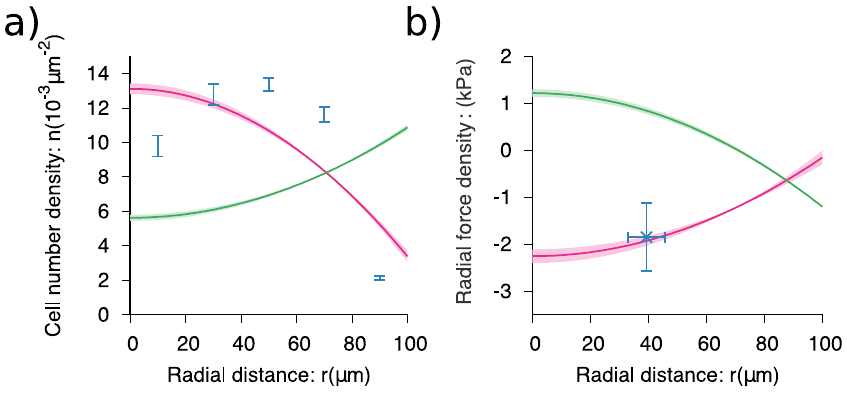}
\caption{(color online) Theoretical fits of steady-state profiles for asters. a) Cell number density $n$, b) radial force density as a function of the radial distance $r$. Averaged experimental profiles (blue, $N=10$ in (a) and $N=3$ in (b)), mean fit in the active-stress dominated (magenta, full lines) and in the traction dominated parameter region (green, dashed lines). Parameters are given in Tab.~\ref{tab:table1}. We used $\zeta''\Delta\mu=0$. Error bars in theoretical fits correspond to std of all values of parameters with $\mathcal{E}<1.1\mathcal{E}_\mathrm{min}$, and experimental curves to sem.}\label{fig4} 
\end{figure}

A further difference between the two regions lies in the mechanics of asters. To study asters in our framework, we set $\psi_0=0$ and kept the other parameter values. In the traction force dominated region, the cell number density increases from the center to the periphery of the confinement, whereas the opposite behavior is obtained in the active stress dominated region, see Fig.~\ref{fig4}a. In our experiments on asters, the cell number density was larger in the center than at the periphery, see Fig.~\ref{fig4}a. This favored the stress gradient dominated regime. 

This result suggests that there is cell compression in the center of asters, which agrees with the force density obtained in our numerical solution~[LongArticle]. To directly determine the nature of the stress in the center of asters, we introduced elastic nonadhesive pillars in the center of the confinement regions~\cite{Guillamat:2020te}. The deformation of these pillars showed that cells were exerting compressive stresses on these objects, see Fig.~\ref{fig4}b. These results exclude the possibility that traction forces dominate over active stress gradients in C2C12 monolayers. 

The combination of the analysis from the polarization and velocity fields in spirals, Fig.~\ref{fig3}, with the cell number density and stresses fields in asters, Fig.~\ref{fig4}, yield the material parameters in the active stress dominated region listed in Tab.~\ref{tab:table1}.

In summary, we have presented a general method for quantifying material parameters of cell monolayers by analyzing isolated topological defects. Small circular confinements allowed us to control the position and topological charge of these defects. Other techniques could be used for this purpose, in particular, by micropatterning the topography of the substrate~\cite{Turiv:2020gq,Endresen:2019uc} and by applying external magnetic fields~\cite{Dua:1996fo}. These methods allow to impose spatiotemporal cell orientation patterns, which in our system were self-organized. Combining these approaches opens a vast range of possibilities to improve our quantitative understanding of cell monolayer mechanics. 

In our experiments, we observed the formation of mounds as the cell number density continued to increase after aster formation~\cite{Guillamat:2020te}. Depending on intracellular properties, cells in the center of the mound either differentiated or the mound continued to grow in height~\cite{Guillamat:2020te}. In the latter case, multicellular protrusions several hundreds $\mu$m tall  grew spontaneously out of the cell monolayer. These observations highlight the importance of defects for understanding biological processes and also provide guiding principles for designing active materials by self-organization.

\acknowledgments{We thank Zena Hadjivasiliou for suggesting the systematic parameter sampling and Jean-Fran\c cois Joanny for discussions.}

\end{document}